\documentstyle[preprint,aps,eqsecnum]{revtex}
\tightenlines

\begin{document}
\draft
\title {TRIVIALITY OF THE QUARK PROPAGATOR IN THE LADDER 
         APPROXIMATION IN QCD}
 
\author{V. Gogohia }

\address{Research Center for Nuclear Physics (RCNP), Osaka University \\ 
         Mihogaoka 10-1, Ibaraki, Osaka 567-0047, Japan  \\
       gogohia@rcnp.osaka-u.ac.jp and gogohia@rmki.kfki.hu}  

 
\maketitle
 
\begin{abstract}
The validity of the ladder approximation (LA) in
QCD and QED in the context of the corresponding Schwinger-Dyson
(SD) equations and Slavnov-Taylor (ST) and
Ward-Takahashi (WT) identities is investigated. In contrast to QED, in QCD because of color degrees of freedom the summation of the
ladder diagrams within the Bethe-Salpeter (BS)
integral equation for the quark-gluon vertex at zero momentum transfer on account of the corresponding ST identity does provide an
addition constraint on the quark SD equation itself. Moreover, the solution of 
the constraint equation requires the full quark propagator should be almost trivial (free-type) one, i. e. only almost trivial quark propagator is allowed in 
the LA to QCD. This triviality results in the fact that the
standard LA ignores the self-interaction between gluons caused by
color charges (non-Abelian character of QCD).
\end{abstract}



\section{Introduction }

 It is well known that the full dynamical information of any quantum
field gauge theory such as QCD is contained in the corresponding quantum
equations of motion, the so-called Schwinger-Dyson (SD)
equations for propagators (lower Green's functions) and
vertices (higher Green's functions) [1]. The Bethe-Salpeter (BS) type integral
equations for higher Green's functions and bound-state amplitudes [2] should be also included into this system. It should be complemented by the corresponding Slavnov-Taylor (ST) identities [1,3,4] which,
in general, relate lower and higher Green's functions to each other. These identities are consequences of the exact gauge invariance and therefore $"are \ exact \ constraints \ on \ any \ solution \ to \ QCD"$ [1]. Precisely
this system of equations can serve as an adequate and
effective tool for the non-perturbative approach to QCD [1,5].
It is important to understand however, that the above mentioned system is an
infinite chain of strongly coupled highly nonlinear integral
equations, so there is no hope for an exact solutions.
For this reason, some truncation scheme is always needed in order
to make these equations tractable for getting physical information
from them.                                                                     

Because of its relative simplicity the most popular is the ladder approximation
(LA) in various forms. In general, it consists of approximating the full vertices by their free perturbative (point-like) counterparts in the corresponding kernels of the above mentioned integral equations. Approximating in addition the full gluon propagator by its free perturbative expression, for example 
in the BS-type integral equation for the vertex, one gets the so-called quenched LA (QLA) scheme which for the quark SD equation became known as "rainbow" approximation. However, very soon it was realized that QLA was too crude for QCD and should be improved in order to incorporate the QCD renormalization group
results for the running coupling constant (asymptotic freedom) into the system
of SD equations. The so-called improved LA (ILA) was proposed [6] (special dependence of the full gluon form factor on its momentum). It makes it possible to 
take into account self-interacting gluon modes (non-Abelian character of QCD) at the level of the full gluon propagator only, so vertices remain intact, i. e.
point-like ones.                   

The main problem of the LA is, of course, its self-consistency. For a long time, it was widely believed that the quark sector could be effectively decoupled
from other sectors in QCD in the LA. For example, it is usually assumed that ghost degrees of freedom are not important for the above mentioned SD system of
equations in this case. At first sight this is so indeed, since neither the quark SD equation nor the equation for the quark-gluon
vertex depends explicitly on ghosts in the LA. However, in QCD as it was mentioned above, these two equations should be complemented by the corresponding ST
identity which explicitly depends on ghost degrees of freedom (even in the LA).
Unlike QED, QCD is much more complicated gauge theory. It contains many different sectors. Making some truncation in one sector, it is necessary to be sure that nothing is going wrong in other sectors since the above mentioned SD 
system of equations may remain strongly coupled even after truncation. The key
elements relating different sectors in QCD are the above mentioned
ST identities, playing thus an important role in the investigation of the problem of self-consistency of any truncation scheme. In other words, any solution to QCD, in particular LA should be compatible with these identities and not vice
versa as it has been emphasized above.        

 Especially important the problem of self-consistency of the LA becomes when  
it is applied to such essentially non-perturbative problems as quark
confinement and dynamical chiral symmetry breaking (DCSB). Its inconsistency for the above mentioned problems was pointed out by Adler [7] who explicitly showed (within precisely the dynamical equations approach) that
a more sophisticated approximation than the LA is needed to
understand quark confinement and DCSB in QCD. The importance of the self-consistent treatment of DCSB within the SD system of equations in the LA was, apparently for the first time, emphasized in Ref. [8]. The next interesting step in
this direction was done quite recently in Ref. [9] where a systematic method was developed for obtaining consistent approximations to the SD and BS equations which maintained the external gauge invariance (by using the Ward-Takahashi 
(WT) identities for the color-singlet, but flavor-nonsiglet vertices). In contrast to the 
above mentioned paper [9], the main purpose of this paper is to investigate the self-consistency of the LA by using the ST identity for the color-nonsiglet, but 
flavor-singlet quark-gluon vertex in QCD (the internal gauge invariance). Precisely this vertex is one of the most important key objects in QCD itself.       
Everybody knows that the LA is bad (for example, it is explicitly gauge-dependent truncation scheme), nevertheless eveybody continues to use it. In some sense
it was justified (see, for example Ref. [10]) since up until  
now there was not an exact criterion to prove or disprove the LA in gauge theories. Here we precisely propose how to formulate this criterion.

In section 2 we remind some well-known things from QED in order to better explain our method. In section 3 we derive the constraint equation and in sections 4
and 5 it is solved without and with ghosts, respectively requiring the full qurk propagator to be almost free one. In section 6 the flavor-nonsinglet but color-singlet axial-vector vertex is investigated. Some general remarks on the non-perturbative renormalization in the Landau gauge is presented in section 7. Our
conclusions are given in section 8.

\section{QED}

   Let us start with some well-known things from QED and consider
first unrenormalized (for simplicity) SD equation
in the LA for the quark (electron) propagator in momentum space

\begin{equation}
S^{-1}(p) = S^{-1}_0(p) - g^2 \int {d^nl\over {(2\pi)^n}} \gamma_\alpha  S(l) \gamma_\beta D_{\alpha \beta}(q),
\end{equation}
where $S^{-1}_0(p) = - i (\hat p - m_0)$ with $m_0$ being the current ("bare")
mass of a single quark (electron) and $D_{\alpha\beta}(q)$ is the full gluon  
(photon) propagator in the arbitrary covariant gauges,                   
                                            
\begin{equation}
D_{\alpha\beta}(q) = - i \left\{ \left[ g_{\alpha\beta} -
{{q_\alpha q_\beta}\over {q^2}} \right]
{1\over {q^2}} d(q^2; \xi) + \xi {{q_\alpha q_\beta}\over {q^4}} \right\}
\end{equation}
and $\xi$ is the gauge-fixing parameter ($\xi=0$, Landau gauge).
Here $q=p-l$ is the transfer momentum   
and we assign the factor $-i g \gamma$ to the point-like ("bare")
veritices with the corresponding Dirac indices.
Differentiating Eq. (2.1) with respect to $p_{\mu}$, one obtains

\begin{eqnarray}
\partial_{\mu} S^{-1}(p) &=& -i \gamma_{\mu} - \partial_{\mu} g^2
\int {d^nl \over {(2\pi)^n}}
\gamma_\alpha  S(l) \gamma_\beta D_{\alpha \beta}(q) \nonumber\\
&=& -i \gamma_{\mu}  + g^2 \int {d^nl \over {(2\pi)^n}}
\gamma_\alpha
[\partial_{\mu} S(l)] \gamma_\beta D_{\alpha \beta}(q),
\end{eqnarray}
up to unimportant total derivative (which is assumed as usual to
vanish at the ends of integration). This is the differential
form of the quark SD equation (2.1) relevant for further
discussion in the next section as well.
 
 Let us consider now the Bethe-Salpeter (BS) integral equation in
the LA for the quark-photon vertex $\Gamma_\mu(p,q)$ ($q=p'-p$)
at zero momentum transfer $q=0$,

\begin{equation}
\Gamma_{\mu} (p,0) = -i \gamma_{\mu} -  g^2
\int {d^nl \over {(2\pi)^n}}
\gamma_\alpha  S(l) \Gamma_{\mu} (l,0) S(l) \gamma_\beta
D_{\alpha \beta}(q).
\end{equation}
The corresponding WT identity in QED,
$q_{\mu}\Gamma_{\mu} (p+q,p) = S^{-1}(p+q) - S^{-1}(p)$,
provides an exact solution for $\Gamma_{\mu} (p,0)$, namely

\begin{equation}
\Gamma_{\mu} (p,0) = \partial_{\mu} S^{-1}(p).
\end{equation}
Substituting it into the Eq.(2.4) and using the obvious identity
$\partial_{\mu} S^{-1}(p) = - S^{-1}(p)[\partial_{\mu} S(p)] S^{-1}(p)$,
one finally obtains

\begin{equation}
\partial_{\mu} S^{-1}(p) = -i \gamma_{\mu}
+ g^2 \int {d^nl \over {(2\pi)^n}} \gamma_\alpha
[\partial_{\mu} S(l)] \gamma_\beta D_{\alpha \beta}(q).
\end{equation}
This is the quark SD equation obtained from the BS integral
equation for the corresponding vertex on account of the WT
identity (2.5). Comparing it with that of Eq. (2.3) (second
line), one immediately concludes that they are identical.
The renormalized version of Eq. (2.3) is obtained
by the multiplication of its inhomogenious term by the quark wave
function renormalization constant $Z_2$, while
the renormalized version of Eq. (2.6) is obtained by the
multiplication of its inhomogenious term by the vertex
renormalization constant $Z_1$. Because of the WT identity (2.5)
these constants are equal to each other ($Z_1=Z_2$) and
again renormalized versions of these equations coincide in QED.               
So the BS equation at zero momentum transfer and on account of
the corresponding WT identity (which is
a consequence of the exact gauge invariance) does not provide an additional
constraint on the quark SD equation in QED.                                    
In other words, the summation of the ladder diagrams within the corresponding  
BS integral equation at
zero momentum transfer on account of the WT identity is consistent with the electron (quark) SD equation itself.

\section{QCD }

The situation in QCD is completely different and is much more complicated because of color and ghost degrees of freedom which precisely provide
an addition constraint on the solution to the quark SD equation in the LA by using the same method as in previous section for QED. It is convenient to begin
our analysis from the quark SD equation in QCD which is just the same as in QED
(2.1) with only trivial replacement $g^2 \rightarrow g^2_F$ because
of color group factors. Here $g^2_F = g^2 C_F$ and $C_F$ is
the eigenvalue of the quadratic
Casimir operator in the fundamental representation (for SU(N), in
general, $C_F = (N^2 - 1)/2N = 4/3,  N=3$). Thus the renormalized version of   
the quark SD equation (2.1) is as follows

\begin{eqnarray}
\tilde{S}^{-1}(p) &=& Z_2 S^{-1}_0(p) + i \tilde{\Sigma}(p) \nonumber\\
          &=& Z_2 S^{-1}_0(p) - \tilde{g}^2_F \int {d^nl \over {(2\pi)^n}}
\gamma_\alpha  \tilde{S}(l) \gamma_\beta \tilde{D}_{\alpha \beta}(q),
\end{eqnarray}
where the renormalized quark propagator and the full gluon propagator (2.2) are
related to the unrenormalized ones as

\begin{eqnarray}
S (p) &=& Z_2 \tilde{S}(p), \nonumber\\                                        
D_{\alpha \beta}(q) &=& Z_V \tilde{D}_{\alpha \beta}(q)
\end{eqnarray}
with the renormalization of the coupling constant is beeing determined
by the corresponding combination of the above and below introduced renormalization constants, $Z's$ [4]. The differential form of the renormalized quark  
SD equation becomes (compare with Eq. (2.3))

\begin{eqnarray}
\partial_{\mu} \tilde{S}^{-1}(p) &=& -Z_2 i \gamma_{\mu} + \partial_{\mu}
i\tilde{\Sigma}(p) \nonumber\\                                              
&=& -Z_2 i \gamma_{\mu}
 + \tilde{g}^2_F \int {d^nl \over {(2\pi)^n}}
\gamma_\alpha [\partial_{\mu} \tilde{S}(l)]
\gamma_\beta \tilde{D}_{\alpha \beta}(q).
\end{eqnarray}

  Let us introduce now the renormalized ST identity [1,3]

\begin{equation}
k_\mu \tilde{\Gamma}^a_\mu(p, k) \left[ Z^{-1}_g +
\tilde{b}(k^2; \xi) \right] = \left[ Z^{-1}_B T^a- \tilde{B}^a(p, k; \xi)
\right] \tilde{S}^{-1}(p+k) -  \tilde{S}^{-1}(p) \left[Z^{-1}_B
T^a - \tilde{B}^a(p, k; \xi) \right],
\end{equation}
where the corresponding renormalized quark-gluon vertex, the ghost self-energy
and the ghost-quark scattering kernel\footnote{For its skeleton expansion see,
for example papers [11,12] and references therein. In the LA only the first term of this expansion survives with the replacement of the corresponding full ghost- and quark-gluon vertices by their free perturbative (point-like) counterparts. However, there is no need to exploit the LA explicit expressions for quantities defined in (3.8) and (3.9) (see below) within our approach.} are related to their unrenormalized counterparts as follows:

\begin{eqnarray}
\Gamma_\mu(p, k) &=& Z^{-1}_1 \tilde{\Gamma}_\mu(p, k), \nonumber\\
b(k^2; \xi) &=& Z_g \tilde{b}(k^2; \xi), \nonumber\\                           
B(p, k; \xi) &=& Z_B \tilde{B}(p, k; \xi).                                     
\end{eqnarray}                                      
For furture aim we have introduced an explicit dependence on a gauge fixing parameter $\xi$ into the ghost degrees of freedom. Let us note that in the
ST identity (3.3) the transfer momentum runs through the ghost degrees of freedom. The following important relation between renormalization constants holds [1]

\begin{equation}
 Z_1 Z_B = Z_2 Z_g.
\end{equation}
It is worthwhile noting that for the sake of furture convenience instead of the ghost field renormalization constant and the "renormalization"
constant $Y$  for the amplitude $H (p, k) = 1
- B(p, k)$, defined in Ref. [1], we have introduced the ghost self-energy remormalization constant $Z_g$ and the "renormalization" constant $Z_B$ for the ghost-quark scattering kernel itself.  

  Differentiation of the renormalized ST identity (3.3) with
respect to $k_\mu$ and then setting $k=0$, yields\footnote{It is assumed that 
possible unphysical kinematic singularities have been already removed from the
vertex by means of Ball and Chiu procedure [13].}

\begin{equation}
\tilde{\Gamma}^a_\mu(p, 0) \left[ Z^{-1}_g + \tilde{b}(0; \xi) \right]
= Z^{-1}_B T^a \partial_\mu \tilde{S}^{-1} (p)
- \tilde{\Pi}^a_\mu (p, 0; \xi)
- \tilde{\Psi}^a_\mu (p; \xi) \tilde{S}^{-1}(p) +
\tilde{S}^{-1}(p) \tilde{\Psi}^a_\mu (p; \xi),
\end{equation}
where we introduced the following notations

\begin{equation}
\tilde{\Psi}^a_\mu(p; \xi) = \left[ {\partial \over \partial k_{\mu}}
\tilde{B}^a (p, k; \xi) \right]_{k = 0}
\end{equation}
and

\begin{equation}
\tilde{\Pi}^a_\mu(p, 0; \xi) = \tilde{B}^a(p, 0; \xi) \partial_\mu
\tilde{S}^{-1} (p).
\end{equation}
Here a few remarks are in order.                                              
Though the regular dependence of the ghost-quark scattering kernel $\tilde{B}^a(p, k; \xi)$ on the ghost self-energy momentum $k$ is preserved by the general Taylor's result [14]     

\begin{equation}
\tilde{B}^a(p, 0; \xi) = \xi F^a (p)= 0 \quad at \quad  \xi=0,
\end{equation}
which shows that this kernel exists at small $k$ in any covariant gauge and vanishes only in the Landau gauge $\xi=0$, nevertheless the regular dependence of the ghost self-energy itself on its momentum is not so obvious. However, here we treat the ghost-self-energy as a regular function 
of its momentum since a singular dependence (which, in principle should not be
excluded $ a \ priori$) requires completely different investigation and is left
for consideration elsewhere. The constraint equation in this case also will 
be completely different from that obtained below.

The renormalized version of the BS-type integral equation for the color non-singlet vertex at zero momentum transfer is

\begin{equation}
\tilde{\Gamma}^a_{\mu} (p, 0) = - Z_1 i \gamma_{\mu} T^a -
\tilde{g}^2 \int {d^nl \over {(2\pi)^n}}
\gamma_\alpha T^b \tilde{S}(l) \tilde{\Gamma}^a_{\mu} (l, 0)
\tilde{S}(l) \gamma_\beta T^b \tilde{D}_{\alpha \beta}(q).
\end{equation}
Substituting (3.7) into the BS-type integral equation (3.11), one obtains

\begin{eqnarray}
Z^{-1}_B T^a \partial_\mu \tilde{S}^{-1} (p) -
\tilde{\Pi}^a_\mu(p, 0; \xi)
&-& \tilde{\Psi}^a_\mu (p; \xi) \tilde{S}^{-1}(p) + \tilde{S}^{-1}(p)
\tilde{\Psi}^a_\mu (p; \xi) = -Z_1 [Z^{-1}_g + \tilde{b}(0; \xi)] i
\gamma_{\mu}T^a
\nonumber\\
&+& \tilde{g}^2 Z^{-1}_B T^b T^a T^b \int {d^nl \over {(2\pi)^n}}
\gamma_\alpha [\partial_{\mu}\tilde{S}(l)]
\gamma_\beta \tilde{D}_{\alpha \beta}(q) \nonumber\\
&+& \tilde{g}^2 \int {d^nl \over {(2\pi)^n}}
\gamma_\alpha T^b \tilde{S}(l) \tilde{\Pi}^a_\mu(l, 0; \xi) \tilde{S}(l)
\gamma_\beta T^b \tilde{D}_{\alpha \beta}(q) \nonumber\\
&-& \tilde{g}^2 \int {d^nl \over {(2\pi)^n}}
\gamma_\alpha T^b \tilde{\Psi}^a_{\mu} (l; \xi) \tilde{S}(l)
\gamma_\beta T^b \tilde{D}_{\alpha \beta}(q)  \nonumber\\
&+& \tilde{g}^2 \int {d^nl \over {(2\pi)^n}}
\gamma_\alpha T^b \tilde{S}(l) \tilde{\Psi}^a_{\mu}(l; \xi)
\gamma_\beta T^b \tilde{D}_{\alpha \beta}(q),
\end{eqnarray}
where the obvious identity,                                                   
$\partial_{\mu} S^{-1}(p) = - S^{-1}(p)[\partial_{\mu} S(p)] S^{-1}(p)$,
has been already used. Let us now use the commutation relation between color matricies $[T^a, T^b] = if_{abc} T^c$, where $f_{abc}$ are the antisymmetric $SU(3)$ structure constants with non-zero values, given for example in Ref. [15].  
Then one obtains,                      

\begin{equation}
T^b T^a T^b = [C_F -{1 \over 2}C_A] T^a,                                       
\end{equation}
where $C_F$ is the above mentioned eigenvalue of the quadratic 
Casimir operator in the fundamental representation while $C_A$ is the same but
in the adjoint representation, $C_A = N$ for $SU(N)$ ($N=3$ for QCD).  
So from this relation and on account of the renormalized version of the differential form of the quark SD equation (3.3) (because of $\tilde{g}^2 C_F = \tilde{g}_F^2$) and using finally an important relation (3.6), one arrives at

\begin{eqnarray}
- \tilde{\Pi}^a_\mu(p, 0; \xi)
&-& \tilde{\Psi}^a_\mu (p; \xi) \tilde{S}^{-1}(p) + \tilde{S}^{-1}(p)
\tilde{\Psi}^a_\mu (p; \xi) = -Z_1 \tilde{b}(0; \xi) i \gamma_{\mu} T^a
\nonumber\\
&-& {1 \over 2} C_A T^a \tilde{g}^2 Z^{-1}_B \int {d^nl \over
{(2\pi)^n}} \gamma_\alpha [\partial_{\mu}\tilde{S}(l)]
\gamma_\beta \tilde{D}_{\alpha \beta}(q) \nonumber\\
&+& \tilde{g}^2 \int {d^nl \over {(2\pi)^n}}
\gamma_\alpha T^b \tilde{S}(l) \tilde{\Pi}^a_\mu(l, 0; \xi) \tilde{S}(l)
\gamma_\beta T^b \tilde{D}_{\alpha \beta}(q) \nonumber\\
&-& \tilde{g}^2 \int {d^nl \over {(2\pi)^n}}
\gamma_\alpha T^b \tilde{\Psi}^a_{\mu} (l; \xi) \tilde{S}(l)
\gamma_\beta T^b \tilde{D}_{\alpha \beta}(q)  \nonumber\\
&+& \tilde{g}^2 \int {d^nl \over {(2\pi)^n}}
\gamma_\alpha T^b \tilde{S}(l) \tilde{\Psi}^a_{\mu} (l; \xi)
\gamma_\beta T^b \tilde{D}_{\alpha \beta}(q).
\end{eqnarray}
This is a renormalized version of the general constraint equation which relates ghost degrees of freedom to those of quark ones in a new manner within the  
LA. In principle, the solution (if any) of the quark SD equation (3.1) itself  
should be compatible with this constarint equation. At first sight it is too complicated integral equation, but nevertheless its analysis is rather simple.
 Here let us note that 
constraint (3.14) is slightly simplified in the Landau gauge $\xi=0$ since the 
composition (3.9) vanishes due to the above mentioned Taylor's general result (3.10).

\section{Solution without ghosts}

As it has been already mentioned in the Introduction, it is widely believed that the ghost degrees of freedom are not important for the LA to covariant gauge QCD. If this is so indeed, let us omit them from the general constraint equation (3.14) by "hand", which is equivalent to consider QED-type ST identity (3.4) 
(apart from the color group generators) from the very beginning. 
Apart from neglecting corrections to the vertex, this is the standard procedure
in the LA to covariant gauge QCD. It has been already done in order to proceed 
from the SD equation for the vertex to its BS-type counterpart (3.11) which 
allows precisely to explicitly sum up ladder contributions (see, for example Ref. [5]). Doing so in (3.14) and on account of Eq. (3.3), one obtains         
          
\begin{equation}
\int {d^nl \over
{(2\pi)^n}} \gamma_\alpha [\partial_{\mu}\tilde{S}(l)]
\gamma_\beta \tilde{D}_{\alpha \beta}(q) = \tilde{g}^{-2}_F  \partial_{\mu}   
i\tilde{\Sigma}(p) = 0,                                              
\end{equation}
which implies general solution is to be $i\tilde{\Sigma}(p)= im_c$,     
where $m_c$ is the constant of integration of the dimension of mass. From the 
quark SD equation (3.1) (first line)

\begin{equation} 
\tilde{S}^{-1}(p) = Z_2 S^{-1}_0(p) + i \tilde{\Sigma}(p)
\end{equation}
with $S_0^{-1}(p) = -i(\hat p -m_0)$ and on account of the above mentioned trivial solution to the quark self-energy, it finally follows 

\begin{equation}
\tilde{S}(p) = { i \tilde{Z}_2 \over \hat p  - \tilde{m} },
\end{equation}
where $\tilde{Z}_2^{-1} = Z_2$ and $\tilde{m} = m_0 + \tilde{Z}_2 m_c$.
Thus the solution of the constraint equation (3.14) with ghosts omitted by "hand" in the LA to covariant gauge QCD requires that the quark propagator should be almost free one apart from the redefinition of the quark mass. In other words, there should 
be $no \ nontrivial$ solution to the quark SD equation in the LA in this case. 
So the main question arises, namely may ghosts cure this problem or not? The final answer 
is, of course, not (see next section). If they would cure this problem it would
mean that the quark propagator (via the general constraint equation (3.14)) would explicitly depend on ghost degrees of freedom. However, this is impossible 
in QCD [1] where ghosts contribute only into the closed loops, so nothing can explicitly depend on them exept of the contributions to the renormalization constants, constants of integration, etc. Precisely the above mentioned problem will be investigated below in more sophisticated fashion.

\section{Solution with ghosts}

In order to investigate the general constraint equation (3.14) in more sophisticated way, let us write down formal BS-type integral equations for the compositions defined in (3.8) and (3.9).
Like the quark-gluon vertex, these objects are vector quantities at zero momentum transfer. They satisfy formally absolutely the same BS-type integral equations in the LA as the above mentioned quark-gluon vertex at zero momentum transfer (3.11). Thus they are the LA equations for (3.8) and (3.9) in the same way as
(3.11) is the LA to the exact BS integral equation for the quark-gluon vertex at zero momentum transfer. The one of the differences is that 
there are no point-like counterparts of these quantites in QCD. But the main difference is that, in contrast to the BS-type integral equation for the quark-gluon vertex (3.11), these equations are completely auxiliarly. They have no independent role, the main purpose to use (or to postulate) them is to show (as was
mentioned above) that nothing explicitly depends on ghost degrees of freedom in
QCD. So in complete analogy with (3.11), the formal homogenious BS-type equation in the LA for the composition (3.9) looks like

\begin{equation}
\tilde{\Pi}^a_\mu(p, 0; \xi) =
- \tilde{g}^2 \int {d^nl \over {(2\pi)^n}}
\gamma_\alpha T^b \tilde{S}(l) \tilde{\Pi}^a_\mu(l, 0; \xi) \tilde{S}(l)
\gamma_\beta T^b \tilde{D}_{\alpha \beta}(q). 
\end{equation}
For the composition defined in (3.8) it is convenient to write down the formal 
BS-type integral equations for its left and right hand side combinations,      
$ \tilde{\Psi}^{a(l)}_\mu (p; \xi) = \tilde{\Psi}^a_\mu (p; \xi) \tilde{S}^{-1}(p)$ and $\tilde{\Psi}^{a(r)}_\mu (p; \xi) = \tilde{S}^{-1}(p)
\tilde{\Psi}^a_\mu (p; \xi)$, respectively

\begin{eqnarray}
\tilde{\Psi}^{a(l)}_\mu (p; \xi) &=& 
- \tilde{g}^2 \int {d^nl \over {(2\pi)^n}}
\gamma_\alpha T^b \tilde{S}(l)\tilde{\Psi}^{a(l)}_{\mu} (l; \xi) \tilde{S}(l)
\gamma_\beta T^b \tilde{D}_{\alpha \beta}(q)  \nonumber\\
\tilde{\Psi}^{a(r)}_\mu (p; \xi) &=& 
- \tilde{g}^2 \int {d^nl \over {(2\pi)^n}}
\gamma_\alpha T^b \tilde{S}(l)\tilde{\Psi}^{a(r)}_{\mu} (l; \xi) \tilde{S}(l)
\gamma_\beta T^b \tilde{D}_{\alpha \beta}(q).
\end{eqnarray}
Substituting (5.1) and (5.2) into the constraint equation (3.14), we are left 
with

\begin{equation}
 - Z_1 \tilde{b}(0; \xi) i \gamma_{\mu} = 
 {1 \over 2} C_A \tilde{g}^2 Z^{-1}_B \int {d^nl \over
{(2\pi)^n}} \gamma_\alpha [\partial_{\mu}\tilde{S}(l)]
\gamma_\beta \tilde{D}_{\alpha \beta}(q).
\end{equation}
From Eq. (3.3), however it follows 

\begin{equation}
\int {d^nl \over
{(2\pi)^n}} \gamma_\alpha [\partial_{\mu}\tilde{S}(l)]
\gamma_\beta \tilde{D}_{\alpha \beta}(q) = \tilde{g}^{-2}_F  \partial_{\mu} i\tilde{\Sigma}(p),                                              
\end{equation}
so Eq. (5.3) by denoting
$\tilde{b}_1(0; \xi)= 2 Z_1 Z_B (C_F/C_A) \tilde{b}(0; \xi)  = 
2 Z_2 Z_g (C_F/C_A) \tilde{b}(0; \xi)$, because of (3.6), becomes
$- i \gamma_{\mu} \tilde{b}_1(0; \xi) =  \partial_{\mu} i\tilde{\Sigma}(p)$. Its general solution is,
$i\tilde{\Sigma}(p)= - i \hat p \tilde{b}_1(0; \xi) + im_c$,     
where $m_c$ is the constant of integration of the dimension of mass. From the 
quark SD equation (4.2) it finally follows 

\begin{equation}
\tilde{S}(p) = { i \tilde{Z}_2 \over \hat p  - \tilde{m} },
\end{equation}
where

\begin{eqnarray}
\tilde{Z}_2^{-1} &=& Z_2 \Big( 1 + 2 Z_g (C_F/C_A) \tilde{b}(0; \xi) \Big), \nonumber\\                                                                       
\tilde{m} &=& \tilde{Z}_2 ( m_c + Z_2 m_0).
\end{eqnarray}
Thus the use of the auxiliarly BS-type integral equations (5.1) and (5.2) allows one to treat ghost degrees of freedom in more sophisticated way than simply omit them by "hand" (as in previous section), but, nevertheless  
the quark propagator in the LA again remains trivial one (5.5) apart from the redefinitions of the quark mass and the quark wave function renormalization constant with the help of ghost self-energy at zero point (compare with (4.3)).  
Nothing explicitly depends on ghosts as it should be indeed in QCD. It
is well known that in the Landau gauge $\xi =0$ the quark wave function is not
renormalized, i. e. $Z_2=1$. Whether the redefined renormalization constant $\tilde{Z}_2$ is also will be equal to unity in this gauge depends on the solution
of the SD equation for the ghost self-energy. This investigation as well as
the solution of the quark SD equation itself in order to determine a possible 
value of the constant of integration $m_c$ is, obviously, beyond the scope of this paper and is left for consideration elsewhere (see also section 7).

\section{QCD. Flavor-nonsinglet axial-vector vertex}

  It is convenient to continue our discussion of the situation in
QCD with the ST identity for the flavor-nonsinglet (but
color-singlet), axial-vector vertex.
It is not complicated by the unknown ghost
contributions. In QED there is no sense to consider axial-vector
vertex because there exists only flavor-singlet channel which,
obviously, will be always complicated by the axial anomaly [16].
Contrast to QED, in QCD there exists flavor-nonsinglet
channel which is free from axial anomaly, but
a new interesting problem occurs with this identity. Also
not loosing generality, let us consider the chiral limit ($m_0 =0$)
of this identity since the nonchiral case, evidently, can not
change conclusions drawn below.

    Let us consider the flavour non-singlet, axial-vector WT
identity in the chiral limit
\begin{equation}
   iq_\mu\Gamma^i_{5\mu}(p+q,p)=\left({\lambda^i \over
 2}\right)
   \{\gamma_5S^{-1}(p)+S^{-1}(p+q)\gamma_5 \},
\end{equation}
where $q=p'-p$ is the momentum transfer (the external momentum) and the quark propagator is given by
 
\begin{equation}
-iS(p)= \hat pA(p^2)+B(p^2),
\end{equation}
so its inverse is expressed as
\begin{equation}
\{-iS(p)\}^{-1} = \hat p \overline A(p^2)-\overline B(p^2),
\end{equation}
with

\begin{eqnarray}
\overline A(p^2) &=& A(p^2)D^{-1}(p^2), \nonumber\\
\overline B(p^2) &=& B(p^2)D^{-1}(p^2), \nonumber\\
D(p^2) &=& p^2 A^2(p^2)-B^2(p^2) .
\end{eqnarray}
 In connection with (6.1), one has to point out that, in
general, $\lambda^i$ is a $SU(N_f)$ flavour matrix  and, in the
massless case, the quark propagator is proportional to the unit
matrix in the flavour space.

 It is well known that if dynamical chiral symmetry breaking
(DCSB) at the quark level, implemented as

\begin{equation}
\bigl\{ S^{-1}(p),  \gamma_5 \bigr \}_+ = i \gamma_5
2 \overline B(p^2) \ne 0
\end{equation}
takes place, then the axial-vector vertex should have a pole
corresponding
to a Goldstone state. Indeed, from (6.5) it follows that
$\Gamma^i_{5 \mu}(p+q,p)$ has a pseudoscalar pole
at $q^2=0$ (dynamical singularity which determines Goldstone
state) if and only if the double of the dynamically
generated quark mass function  $2 \overline B (p^2)$ in (6.5)
is nonzero and vice versa. In order to self-consistently
untangle regular and pole parts in (6.1), it is necessary to
perform a procedure described in detail in our paper [12]. It is a
similar to that of the above mentioned Ball and Chiu procedure
[13] to remove kinematic (unphysical) singularities from the
vertex. As a result of this, dependence on
the two arbitrary form factors occurs in the regular part of this
vertex at zero momentum transfer $q=0$, namely 

\begin{equation}
 \Gamma^{iR}_{5 \mu}(p,p) = \left({\lambda^i \over 2} \right)
[- i {\partial_{ \mu} S^{-1}(p)} + \Delta_{ \mu}(p)] \gamma_5,
\end{equation}
where

\begin{equation}
\Delta_{ \mu}(p) = \gamma_{ \mu} R_6(p^2)
 - \hat p \gamma_{ \mu} R_{11}(p^2).
\end{equation}
The corresponding BS integral equation becomes equation for
these form factors and therefore no statement about justification
of the LA can be deduced. Indeed, let us show this explicitly. 

   The BS integral equation for the regular part of the
axial-vector vertex at zero momentum transfer is (here and below in this section again $q=p-l$)

\begin{equation}
\Gamma^{iR}_{5 \mu}(p,p) = \left({\lambda^i \over 2} \right)
\gamma_5 \gamma_{\mu} - g^2_F \int {d^nl\over {(2\pi)^n}}
\gamma_\alpha  S(l) \Gamma^{iR}_{5 \mu} (l,l) S(l) \gamma_\beta
D_{\alpha \beta}(q),
\end{equation}
where summation over color indices is already done, i. e.
$g^2_F = g^2 C_F$ (see Eq. (3.1)).
 
 Substituting relations (6.6-6.7) into this equation and on
account of the above used identity, one obtains

\begin{eqnarray}
[{\partial_{ \mu} S^{-1}(p)} + i\Delta_{\mu}(p)] \gamma_5
&=& i \gamma_5 \gamma_{\mu} \nonumber\\
&+& g^2_F \int {d^nl\over {(2\pi)^n}}
\gamma_\alpha [{\partial_{\mu} S(l)}] S^{-1} (l) \gamma_5
S(l) \gamma_\beta D_{\alpha \beta}(q) \nonumber\\
&-& i g^2_F \int {d^nl\over {(2\pi)^n}}
\gamma_\alpha S(l) \Delta_{\mu} (l) \gamma_5 S(l)
\gamma_\beta D_{\alpha \beta}(q).
\end{eqnarray}

Let us now introduce the following notation

\begin{equation}
S^{-1}(p) \gamma_5 S(p) = - \gamma_5 + \Delta (p) \gamma_5.
\end{equation}
where, obviously $\Delta(p)$ can be given in terms of the quark propagator     
form factors (6.4). 
Multiplication of Eq. (6.9) by $\gamma_5$ from right and on
account of (6.10), yields 
\begin{eqnarray}
{\partial_{ \mu} S^{-1}(p)} + i \Delta_{\mu}(p)
&=& - i \gamma_{\mu} \nonumber\\
&+& g^2_F \int {d^nl\over {(2\pi)^n}}
\gamma_\alpha [{\partial_{\mu} S(l)}]
\gamma_\beta D_{\alpha \beta}(q) \nonumber\\
&-& g^2_F \int {d^nl\over {(2\pi)^n}}
\gamma_\alpha [{\partial_{\mu} S(l)}] \Delta (l)
\gamma_\beta D_{\alpha \beta}(q)  \nonumber\\
&-& i g^2_F \int {d^nl\over {(2\pi)^n}}
\gamma_\alpha S(l) \Delta_{\mu}(l) \gamma_5 S(l)
\gamma_\beta \gamma_5 D_{\alpha \beta}(q).
\end{eqnarray}
Using futher the QCD quark SD equation (3.3), one finally obtains

\begin{eqnarray}
i \Delta_{\mu}(p) =
&-& g^2_F \int {d^nl\over {(2\pi)^n}}
\gamma_\alpha [{\partial_{\mu} S(l)}] \Delta (l)
\gamma_\beta D_{\alpha \beta}(q) \nonumber\\
&-& g^2_F \int {d^nl\over {(2\pi)^n}}
\gamma_\alpha S(l) \Delta_{\mu} (l) \gamma_5 S(l) 
\gamma_\beta \gamma_5 D_{\alpha \beta}(q),
\end{eqnarray}
where, recalling $\Delta_{\mu}(p)$ is given in (6.7). 
This is the inhomogenious BS integral equation for unknown form
factors $R_6$ and $R_{11}$, so, as was mentioned above, no
statement about justification of the LA can be deduced. Evidently,
this is a result of the fact that the exact expression for the
longitudinal part of color-singlet but flavor-nonsinglet
axial-vector ST identity is known up
to two arbitrary form factor even at zero momentum transfer. Also
obviously, that one can derive the similar BS equation for these
form factors in some other approximation for the BS scattering
kernel and again no statement can be achived in favour of this or
another truncation scheme.

\section{Remarks on the non-perturbative renormalization in the Landau gauge}

In this section we are not restricted to the LA and are going to make a few general remarks concerning the non-perturbative renormalization in the Landau
gauge. It is well-known that in QCD the renormalization program at
zero point is greatly simplified in the Landau gauge. The ghost
degrees of freedom, such as the ghost-gluon vertex and
the ghost-quark
scattering kernel, do not renormalize at zero point in the Landau
gauge. Their renormalization constants (defined at zero
point) become trivial unites in this gauge. Summary of
these results one can find in Ref. [1]. This fails if the
renormalization is performed at nonzero euclidean mass scale
parameter (($-\mu^2$), renormalization point) because of the
possible infrared (IR) singularities due to exchange of massless
gluons. However, as was mentioned above, the IR singular behaviour of the ghost
self-energy requires special treatment and will be
considered elsewhere.\footnote{The treatment of ghost degrees of freedom in
the Landau gauge in Ref. [17] is not completely self-consistent. It was not investigated whether the enhancement of the ghost propagator in the IR was compatible with Taylor's general result (3.10) for the ghost-quark scattering kernel which was simply omitted by hand in their truncation scheme.}

Let us consider now the exact ghost propagator and its self-energy which are   
related to each other as follows

\begin{equation}
G(k) = {i \over k^2 [1 + b(k^2; \xi)] },
\end{equation}
where, for simplicity's sake, we suppressed the color group indicies.
Introducing the renormalized ghost propagator

\begin{equation}
G(k) = Z_G \tilde{G}(k),
\end{equation}
from (7.1-7.2) and (3.5), one obtains

\begin{equation}
 Z_G = Z^{-1}_g
\end{equation}
and

\begin{equation}
\tilde{G} (k) = {i \over k^2 [1 + b^R(k^2; \xi)] },
\end{equation}
where we define as usual (see, for example, Ref. [4])

\begin{equation}
b^R(k^2; \xi)= \tilde{b}(k^2; \xi) - \tilde{b}(0; \xi)
\end{equation}
and

\begin{equation}
Z^{-1}_g + \tilde{b}(0; \xi) = 1.
\end{equation}

  In a similar way let us define the regularized quark-ghost scattering amplitude (3.5) as follows

\begin{equation}
B^R(p, k; \xi)= \tilde{B}(p, k; \xi) - \tilde{B}(0, 0; \xi)
\end{equation}
and

\begin{equation}
Z^{-1}_B - \tilde{B}(0, 0; \xi) = 1,
\end{equation}
so that

\begin{equation}
Z^{-1}_B - \tilde{B}(p, k; \xi) = 1 - B^R(p, k; \xi)
\end{equation}
From Taylor's general result (3.10) and on account of (7.8)
then it follows that in the Landau gauge $\xi=0$, one has

\begin{equation}
Z_B = 1,
\end{equation}
and consequently from (3.6) it follows

\begin{equation}
Z_1 = Z_2  Z_g.
\end{equation}
It is worthwhile noting that the
renormalization constant $Z_B$ coincides with the renormalization
constant $Y$ defined in Ref. [1] for the amplitude $H (p, q) = 1
- B(p, q)$, i.e. $Z_B \equiv Y$ in the Landau gauge $\xi=0$ at least. 

 Let us now briefly discuss a possible interesting
feature of the renormalization at zero point in the Landau gauge.
Indeed, from (7.11) it follows that if the ghost self-energy
renormalization constant $Z_g$ (defined at zero point) would may
equal to unity as $Z_B$ (7.10) then the QED-type relation
between the QCD renormalization constants (defined at zero in the
Landau gauge) will be recovered, namely (7.11) would become

\begin{equation}
Z_1 = Z_2.
\end{equation}
But from (7.6) then it follows that the ghost self-energy at
zero point in the Landau gauge should become zero, i.e.
$\tilde{b}(0; \xi=0)=0$. This would be possible only within the
nonperturbative treatment of the corresponding ghost self-energy
SD equation. The problem is that the
perturbative calculation of this renormalization constant
within the LA 
to the above mentioned SD equation suffers from the IR
singularities due to exchange of massless gluons. In this case
this would mean that the ghost propagator in the Landau gauge does
not renormalize at zero point as well as the ghost-gluon vertex
and ghost-quark scattering kernel [1]. This would
be an example of the nonperturbative finite calculation of the
corresponding renormalization constant and deserves furture
investigation elsewhere.

 A few concluding remarks are in order. The SD ghost self-energy equation is
complicated nonlinear integral equation relating three
various Green's functions to each other. There are no doubts that
this equation may have regular as well as singular solutions even in the LA.
However, it is necessary to clearly
understand that the disappearance of the ghost-self energy at zero
point in the Landau gauge is completely nonperturbative effect.
One can not obtain this result calculating only finite number
of perturbative corrections. Only summing up infinite number of
perturbative diagrams within the corresponding SD equation (i.e.
going beyond the perturbation theory), one may hope to achieve
this goal. Moreover, it depends heavily on the nature of the
approximation scheme. i.e. not any approximation scheme may
satisfy this exact result. In other words, the above mentioned possible zero ghost self-energy in the Landau gauge is an
example of the nonperturbative finite calculation of the
corresponding renormalization constants at zero point since the
perturbative renormalization at this point suffers from the IR
singularities due to exhange of massless gluons.
Precisely for this reason, the perturbative renormalization is
always performed at nonzero space-time momentum $k^2 =-\mu^2$ in
order to avoid (not to solve, of course) these difficulties.
At the same time, as it seems to us, the
nonperturbative renormalization will make it possible to perform finite
calculations of the corresponding renormalization constants at zero
point without any problems, i. e. only beyond the perturbation theory
one is able to renormalize theory at zero point at least in the Landau gauge.
This emphasizes once more the special role of the Landau gauge in QCD.

\section{Conclusions}

 In summary, we have formulated the method how to prove or disprove the LA in 
gauge theories such as QED and QCD. In QED the summation of the ladder diagrams
within the BS integral equation for the quark-photon vertex at zero momentum
transfer on account of the corresponding WT identity does not provide an addition constraint on the solution to the quark SD equation itself. In other words, 
there is no criterion to prove or disprove the use of the LA in QED. In contrast to this, in QCD because of color degrees of freedom the summation of the ladder diagrams within the BS integral equation for the quark-gluon vertex at zero 
momentum transfer on account of the corresponding ST identity $does$ 
provide an addition constraint on the solution to the quark SD equation itself.
Moreover, the solution of the constraint equation (3.14) requires that the full
quark propagator should be almost trivial (free-type) one (4.3) or (5.5), i. e.
there is $no \ nontrivial$ quark propagator in QCD in the LA. In other words, there is no running quark mass in the LA to QCD as well. This is our main result. It does not depend on how one treats ghost degrees of freedom in the 
LA to covariant gauge QCD, omitting them by "hand" in the general constraint equation (3.14) or
in more sophisticated fashion by using auxiliarly BS-type integral equations (5.1) and (5.2) there. Let us underline once more, that the existence of the constraint equation (3.14) in the LA to QCD in any gauge is only due to color degrees of freedom and not ghosts. Let us make one thing perfectly clear. The quark SD equation, remainning a nonlinear integral equation even in the LA, may 
have formally a number of non-trivial, approximate solutions. However, non of 
them (analytical or numerical) will satisfy the constraint which comes from the
exact ST identity, so all results based on these solutions should be reconsidered.                   
                
In noncovariant gauges [18] ghosts do not contribute from the very beginning, 
i. e. the ST identity (3.4) becomes of the QED-type apart from the color degrees of freedom. This means that the general constraint equation (3.14) automatically becomes (4.1). Thus from our consideration, one easily can conclude that the 
quark propagator remains trivial one (4.3) in the LA to noncovariant gauge QCD as well (let us remind that we neither use the explicit expression for the full gluon propagator (2.2) nor specify the full gluon form factor there).   
 The physical reason of the discovered triviality of the quark propagator in the LA is, of course, that the LA in the quark sector ignores the self-interactions of gluons caused by color charges (non-abelian character of QCD).        
On the other hand, there is no doubt that the LA (especially in the ILA form) 
is legitimated to use for the ultraviolet (UV) region in QCD because of asymptotic freedom. The problem how to correctly go beyond the LA in order to describethe IR region in QCD becomes inevitably important. It is perfectly clear now that we can not use the LA (in QLA or even in ILA forms (since as 
it was mentioned above we did not specify the full gluon form factor)) in the 
entire energy-momentum range in QCD. The ILA itself is not sufficient to approximate the IR region in QCD. The nontrivial generalization of the point-like vertices is needed for this purpose. We have investigated the self-consistency of 
the LA in QCD. It is almost obvious that any truncation scheme, for example such as planar, $1/N_c$ limit, etc maight and should be investigated in the same way. As it is emphasized in our paper, this is important in gauge theories.

 We didnot discuss the flavor-nonsinglet (but color-singlet)
vector vertex to which external gauge fields such as $W$ and
$Z$ bosons could couple. Evidently,  the situation for this vertex
is completely analogous to that of QED. In other words, contrast
to the axial-vector vertex considered in Section 6, here the use
of the LA is justified as in QED. First it was explicitly shown
in Ref. [9] though in the different way. However, in the light
of our result a
general question is in order. How is it possible to use nontrivial
solutions to the quark SD equation in the LA in the QCD
boun-state problems when one knows that it becomes trivial
(almost free) in the scattering problems?  At long last, the quark
SD equation is the same for both problems. At least in one case,
namely in
the constituent quark model (CQM) the use of the LA is justified
for both problems. Indeed, from a QCD theoretical field point of
view, apart from other simplifications, the CQM mainly is nothing
but an approximation of the full quark
by the constituent quark propagator (which coincedes with (5.5) due to replacement $\tilde{m} \rightarrow m_q$, where $m_q$ is the constituent quark mass) and the full constituent quark-gluon vertex by point-like one. From our results then it follows that in CQM
the quark propagator and quark-gluon vertex is in one-to-one correspondence with each other. Precisely, this self-consistent relation between them explains the success of this calculation scheme despite its simplistic structure.

In conclusion, let us make a few technical remarks. In QED expansion in powers
of the external momentum makes no great sense since there is no stable bound-states (positronium is unstable). In contrast to this, in QCD the same expansion makes precisely great sense because of the existence of the Goldstone sector
there. Many important physical quantities such as scattering lenght, pion charge radius, etc are defined as coefficients of the pion form factor expansion in powers of external momentum. The chiral perturbation theory [19] and its counterpart at the fundamental quark level [20] are also based on these expansions. Thus in QCD zero external momentum transfer has a physical  
meaning as relating directly to various physical observables.

The author would like to thank S.Adler for correspondence and A.T.Filippov and 
A.A. Slavnov for interesting remarks. He is also grateful
to F.Schoberl and W.Lucha for critical discussions and useful remarks on first 
stage of this investigation.


\end{document}